\documentclass[reqno,10pt,centertags]{amsart}
\usepackage{amssymb,upref,graphicx,lastpage}

\allowdisplaybreaks \numberwithin{equation}{section}

\usepackage{graphicx,lastpage}
\makeatletter
\def\@maketitle@hook
  {\hbox to \textwidth
    {\valign{\vss##\vss\cr
      \noalign{\hfil}
      \hsize=0.7\hsize
      \vbox{\parindent0pt \raggedleft

         \medskip
         \sffamily
         Transactions of The\\ Royal Norwegian Society\\ of Sciences and
         Letters,\\
         (\textit{Trans.\ R.\ Norw.\  Soc.\  Sci.\ Lett.} 20xx(x) x--x)\par}\cr}}
  \vskip 42pt \relax}
\makeatother

\begin{document}

\title[]{What do experiments in optics tell us about photon momentum in media?}

\author[]{Iver Brevik}
\address{Department of Energy and Process Engineering, Norwegian University of Science and Technology, N-7491 Trondheim, Norway}
\email{iver.h.brevik@ntnu.no}
\urladdr{http://folk.ntnu.no/iverhb}


\dedicatory{Dedicated  to Johan Skule H{\o}ye}
\date{\today}

\begin{abstract}
In order to get some insight in the intricacies of the Abraham-Minkowski energy-momentum problem in phenomenological electrodynamics, we briefly analyze eight different experimental situations in radiation optics. Six of the experiments are already existing, while the remaining two are suggestions for future endeavors. Among the first six, we distinguish between three which are incapable of informing about photon momentum in a medium, and the remaining three which are able to give us useful information. Our general conclusion is that the Abraham-Minkowski "problem" is essentially a matter of convenience. In actual physical situations it is the Minkowski expression which is most natural alternative to employ. Also, in a canonical context, it is the Minkowski energy-momentum tensor  which is the most convenient alternative to work with, as this tensor is divergence-free causing the total radiation momentum and energy to make up a four-vector, although notably a spacelike one.
\end{abstract}

\maketitle{Photon momentum in  media}
\section{Introduction}

The 100-years-old Abraham-Minkowski problem in the electromagnetic theory for a continuous medium has attracted considerable interest in the recent past (a few reviews and recent research articles can be found in Refs.~\cite{moller72,brevik79,loudon04,baxter10,barnett10,milonni10,mansuripur10,griffiths12,rikken12,ravndal12,webb13}; especially the Resource Letter of Griffiths \cite{griffiths12} contains a wealth of references). The issue is essentially the following: what is the correct (or, more appropriately, the most convenient) expression for the momentum of a radiation field in a medium? Equivalently, one may ask: what is the mechanical force from a field on the medium? One might think that the problem could be solvable merely by a measurement of the force straight away, but it has turned out that such a measurement is far from trivial; this being due, of course, to the smallness of the force under normal circumstances in optics.

There are two different strategies one can follow when trying to figure out the photon momentum problem in matter. One is to start from the microscopic electromagnetic theory, and then make the separation into an electromagnetic part and a mechanical part in such a way the "correctness" or, more appropriately, the convenience, of the division into the two subsystems follows. It is clear from the beginning that such an approach will be encountered with ambiguity problems. The other strategy, which in our opinion is a better one, is to look for the experiments, and derive from that information what is the best alternative for describing the physics. What may be looked upon as a drawback of the latter approach as far as fundamental detailed insight is concerned, may be outweighted with respect to simplicity and usefulness.

Some years ago we presented a review over various experiments in electromagnetism and their importance for the energy-momentum tensor \cite{brevik79}. Here, we will to some extent update that review, limiting ourselves to optical phenomena. The following section sketches some fundamentals of the classical electromagnetic theory in continuous matter, and the subsequent sections briefly analyze some existing experiments and their importance for the momentum problem. We also present a couple of proposals for experiments that might be of interest to carry out in the future.

\section{Extracts from the general formalism}

Consider an electromagnetic field within an isotropic dielectric medium whose real refractive index is $n=\sqrt{\varepsilon \mu}$ in the rest system. In an arbitrary inertial system where the medium as a whole has a uniform four-velocity $V_\mu$, the Minkowski energy-momentum tensor for the field is
\begin{equation}
S_{\mu\nu}^{\rm M}=F_{\mu\alpha}H_{\nu \alpha}-\frac{1}{2}\delta_{\mu\nu}F_{\alpha\beta} H_{\alpha \beta}, \label{1}
\end{equation}
where in standard notation $F_{4k}=(i/c)E_k, H_{4k}=(i/c)D_k, F_{ik}=B_l$ (cycl), $H_{ik}=H_l$ (cycl). We employ the conventions $x_\mu=(x_i, x_4=ict)$, implying $V_\mu V_\mu=-c^2$ (these are as in M{\o}ller's book \cite{moller72}, except from our use of SI units).  The spatial components of $S_{\mu\nu}^{\rm M}$ are
\begin{equation}
S_{ik}^{\rm M}=-E_iD_k-H_iB_k+\frac{1}{2}\delta_{ik}({\bf E\cdot D+H\cdot B}). \label{2}
\end{equation}
The Poynting vector ${\bf S}^{\rm M}$ and the electromagnetic momentum density ${\bf g}^{\rm M}$ are determined via $S_{4k}^{\rm M}=(i/c)S_k^{\rm M}$ and $S_{k4}^{\rm M}=icg_k^{\rm M}$ to give
\begin{equation}
{\bf S}^{\rm M}=\bf E\times H, \label{3}
\end{equation}
\begin{equation}
{\bf g}^{\rm M}=\bf D\times B. \label{4}
\end{equation}
The corresponding energy density is
\begin{equation}
w^{\rm M}=\frac{1}{2}\bf (E\cdot D+H\cdot B). \label{5}
\end{equation}
If we formally put $\varepsilon_0=\mu_0=1$ and write the constitutive relations in the rest system  ($V_i=0$)  in the form ${\bf D}=\varepsilon \bf E$, ${\bf B}= \mu \bf H$, we can conveniently write them in a general inertial frame  as
\begin{equation}
\mu H_{\mu\nu}=F_{\mu\nu}-\frac{n^2-1}{c^2}\left(F_{\mu\alpha}V_\nu-F_{\nu\alpha}V_\mu\right)V_\alpha. \label{6}
\end{equation}
In the following we shall however write the constitutive relations as ${\bf D}=\varepsilon_0\varepsilon \bf E$, ${\bf B}=\mu_0 \mu \bf H$, letting $\varepsilon$ and $\mu$ be nondimensional.

 In the rest system we get for the Minkowski force density
\begin{equation}
{\bf f}^{\rm M}=\rho {\bf E}+{\bf J\times B}-\frac{1}{2}\varepsilon_0 E^2{\bf \nabla}\varepsilon -\frac{1}{2}\mu_0H^2{\bf \nabla}\mu,
\quad f_4^{\rm M}=0, \label{7}
\end{equation}
where $\rho$ is the charge density and $\bf J$ the current density. Electrostrictive effects are here omitted.

As a general remark, the Minkowski tensor  turns out to be convenient for the  description of  optical phenomena. For instance, if $\rho ={\bf J}=0$, and if the medium is homogeneous so that ${\bf \nabla}\varepsilon={\bf \nabla}\mu=0$, one has $\partial_\nu S_{\mu\nu}^{\rm M}=0$, implying that the components of total field momentum and total energy make up a four-vector. This is a nontrivial point, since the electromagnetic field in a medium is a non-closed system (the other sub-system being the mechanical counterpart). The Minkowski  four-momentum is even spacelike, meaning that the field energy can become negative in certain inertial systems, as exemplified in the theory of the Cherenkov effect considered in the system where the radiating particle is at rest.

The Abraham energy-momentum tensor $S_{\mu\nu}^{\rm A}$ will be considered in the rest system only. As we have assumed isotropy, the stress tensor components as  well as the Poynting vector  and the energy density are just the same as in the Minkowski case,
\begin{equation}
S_{ik}^{\rm A}=S_{ik}^{\rm M}, \quad {\bf S}^{\rm A}={\bf S}^{\rm M}, \quad w^{\rm A}=w^{\rm M}. \label{8}
\end{equation}
We will denote these variables simply as $\{ S_{ik},{\bf S}, w\}$.

The difference occurs only in the expression for the momentum density,
\begin{equation}
{\bf g}^{\rm A}=\frac{1}{c^2}\bf E\times H. \label{9}
\end{equation}
This arises because of the required symmetry of all the Abraham energy-momentum components. It  implies that this tensor satisfies the so-called Planck's principle of inertia of energy,
\begin{equation}
{\bf g}^{\rm A}={\bf S}/c^2, \label{10}
\end{equation}
in contrast to what holds in the Minkowski case,
\begin{equation}
{\bf g}^{\rm M}=n^2{\bf S}/c^2. \label{11}
\end{equation}
The non-symmetry of the Minkowski tensor with respect to the fourth row and fourth column in the rest system will in other systems be carried over to the spatial tensor components, this being of importance for the calculation of electromagnetic torques on dielectric media in motion.

 We assume now  that $\rho={\bf J}=0$, and take the medium to be nonmagnetic. The Abraham force density then becomes, when spatial inhomogeneity is allowed for,
\begin{equation}
{\bf f}^{\rm A}=-\frac{1}{2}\varepsilon_0E^2{\bf \nabla}n^2+\frac{n^2-1}{c^2}\frac{\partial}{\partial t}(\bf E\times H). \label{12}
\end{equation}
Here the first term is common for the Abraham and Minkowski alternatives and will henceforth be referred to as ${\bf f}^{\rm AM}$:
\begin{equation}
{\bf f}^{\rm AM}= -\frac{1}{2}\varepsilon_0E^2{\bf \nabla}n^2. \label{13}
\end{equation}
 For a homogeneous medium this kind of force occurs in the boundary layers only. The second term in Eq.~(\ref{12}) is the so-called Abraham term. We will call it ${\bf f}^{\rm Aterm}$. Thus
\begin{equation}
{\bf f}^{\rm Aterm}= \frac{n^2-1}{c^2}\frac{\partial}{\partial t}(\bf E\times H). \label{14}
\end{equation}
Under usual circumstances, in an optical stationary field, this term will fluctuate out.

It may be mentioned that the expression (\ref{12}) for the force density agrees with Ginzburg \cite{ginzburg89} as well as with Landau and Lifshitz \cite{landau84}.

We  consider now  examples, briefly covering in the next section some cases from which {\it no} information about photon momentum in matter can be extracted.

\section{Experiments not testing photon momentum}

\subsection{ Radiation pressure on a flat liquid interface}

It is natural to begin with the classic radiation pressure experiment of Ashkin and Dziedzic \cite{ashkin73} (cf. also the useful reprint volume edited by Ashkin \cite{ashkin08}). Focused light was sent from above towards an air-water flat surface and an outward bulge of the order of 1$~\mu$m was observed. The laser was a pulsed single transverse mode doubled Nd:YAG operating at a free space wavelength of $0.53~\mu$m. The pulse repetition frequency was 3 kW (sufficiently low to avoid nonlinear effects), and the duration of each pulse was about 60 ns. The radius of the waist was small, about $4.5~\mu$m. The reason for the low value of the elevation is the large air-water surface tension. The theory of this experiment can be found in Refs.~\cite{brevik79} and \cite{lai76}.

The central point in our context is that this experiment involves only the Abraham-Minkowski force ${\bf f}^{\rm AM}$, acting in the surface. The Abraham term ${\bf f}^{\rm Aterm}$ fluctuates out when averaged over an optical period. The Abraham-Minkowski momentum problem is not involved at all.

It is here worthwhile to point out that the magnitude of the elevation can be enhanced very much by working with a two-fluid system near the critical point. The surface tension can in that way be reduced to a millionth of that of an air-water surface. The surface becomes accordingly sensitive even to small radiation-induced stresses. A few  references  to this kind of experiment are \cite{casner03,hallanger05,birkeland08,wunenburger11}, with further references therein.  It turned out that the laser light, originally falling onto the interface when it was flat, was able to produce a deflection of about 70$~\mu$m at maximum laser power, which was about 1 watt. A recent theoretical treatment of this experiment can be found also in Ref.~\cite{aanensen13}.

\subsection{Radiation pressure on a curved liquid interface}

There exist striking experiments of Zhang and Chang \cite{zhang88}, measuring the deformation of a micro water droplet when illuminated by a laser pulse. Typical pulse energies were about 100 mJ. This duration is very short compared to the hydrodynamic response time of the droplet. The pulse transfers an impulse to the droplet and gives rise to oscillations of the surface, gradually dying out because of viscosity. The main surface deformations occur at the rear, since the droplet acts as a lens. For a 100 mJ pulse, if $a$ is the radius and $h$ the surface elevation, the experiment gave $h/a=0.02$ at the front and $h/a=0.3$ at the rear.

Again the sole agent responsible for the surface elevation is the force ${\bf f}^{\rm AM}$ acting in the boundary layer. The photon momentum in the interior does not come into play. Theoretical papers discussing this effect of microsphere distortions can be found in Refs.~\cite{lai89,brevik99,ellingsen13}.

\subsection{Transverse displacement of a fiber when illuminated by a laser }

Recent years have seen an increased interest in radiation forces in optics: Optical tweezers, atom traps, optical manipulation of soft microparticles, only to name a few. The experiment of She {\it et al.} \cite{she08} shows a particular example of this sort, namely that a vertically hanging silica glass fiber can be exposed to a transverse optical force, and thus a sideways displacement, when transmitted by a laser beam. The beam may be pulsed, or it may be continuous. The authors even claimed that they had solved the Abraham-Minkowski momentum problem by demonstrating that the fiber was carrying the Abraham momentum.

Assume the fiber is fixed at the upper end, and that the lower end is free. There will be a downward directed impulse imparted to the fiber at the lower end when the pulse leaves. This force is unambiguous.  Take the pulse to be short   with energy $\mathcal{H}$=2.7 mJ. When this pulse falls  upon the fiber's entrance surface,  if we take the flux to be  10 mW and the pulse duration to be 270 ms, we get $\mathcal{H}$=2.7 mJ resulting a downward impulse of 4.5 pN$\cdot$s if the refractive index is $n=1.5$ \cite{brevik09}. But where does the {\it transverse} force come from? In our opinion it is most unlikely that it is related to electromagnetic momentum in the fiber. Rather, we suggest that the most natural explanations are  either, (i) that there is  a mechanical imbalance in the fiber resulting from the drawing process, thus resulting in nonperfect azimuthal symmetry, or (ii) that the lower end of the fiber is not cut exactly orthogonal to the length direction. In the latter case, the surface force at the lower end, necessarily directed orthogonally to the surface, will get a sideways component. In either case, it is again the  Abraham-Minkowski force ${\bf f}^{\rm AM}$ which is the actual observable force.

We mention that Manuripur has also raised scepticism with respect to the original interpretation of this experiment \cite{mansuripur09}.

\section{Experiments testing electromagnetic momentum}

We now turn to examples of experiments, real or on the proposal stage, which are in principle able to give information about the photon momentum in a radiation field.

\subsection{Radiation pressure experiment on an immersed mirror}

One of the most important series of experiments in our context are those  of Jones {\it et al.}. In 1951, Jones \cite{jones51} reported in a short note how the pressure exerted by a radiation field on an immersed mirror in a liquid depended on the refractive index $n$, and in 1954 Jones and Richards \cite{jones54} gave an extensive treatment of their experiment (cf. also \cite{jones78}).  Later, Jones and Leslie \cite{jones78a} published a detailed report repeating their earlier experiment with a more than tenfold improvement in accuracy and also  generalized the experiment to cover  the off-normal case.  (The  book by Jones written much later, in which he summarizes his scientific activities, is also highly recommended \cite{jonesbook}.)

Restricting ourselves to the normally incident case, the basic setting of this experiment was the following: A ray of light emitted from a 30 W tungsten filament lamp was sent through a glass window and reflected in the opposite direction from a metallic wall of rhodium-plated silver. (Actually, two rays of light were used, falling asymmetrically on a vane, and the vane was mounted on a torsional pendulum.)  Testing various liquids, including water and benzene, the radiation pressure on the wall was found to vary in proportion to the refractive index $n$ of the liquid. The authors carefully took into account errors arising from various effects, such as absorption in the liquid, multiple reflections at the vane and the window, and dependence of the reflection coefficient $R$ on the index $n$. Unwanted effects from convective forces in the liquid were eliminated by means of a chopping technique. The mechanical torque measured on the vane was very small, of the order of $10^{-13}$ Nm.

As discussed in \cite{brevik79} this experiment can be described in three different ways. We let $x$ be the direction of propagation and $\sigma_x$ the normal surface stress on the wall.

(1) Probably the most natural way to calculate $\sigma_x$ is to identify it with the momentum flux component of the energy-momentum tensor,
\begin{equation}
\sigma_x=S_{xx}=\frac{n}{c}(1+R)S_x^{(i)}, \label{15}
\end{equation}
where $S_{x}^{(i)}$ is the Poynting vector for the  incident plane wave in the liquid.

Viewed from this angle, it is the momentum {\it flux}, rather than the momentum itself, which is the chief agent behind the pressure. The Abraham-Minkowski problem does not come into play explicitly. Dividing by the vacuum (air) pressure $\sigma_0=(1/c)(1+R_0)(S_{x}^{(i)})_0$, and setting $R_0=R$ and $ (S_{x}^{(i)})_0=S_{x}^{(i)}$ (the same Poynting vector in the two cases), we see that the simple proportionality $\sigma_x/\sigma_0=n$ results, as observed in the experiment.

(2) Alternatively, the radiation pressure $\sigma_x$ can be found by recognizing that the Lorentz force density in the metal (assumed nonmagnetic) is directed in the $x$ direction and is equal to $\mu_0\sigma E_y H_z$ (in the real representation) if $\sigma$ is the conductivity and the incident field is taken to be polarized in the $y$ direction. We can evaluate $\sigma_x$ as the integral of this expression from $x=0$ to infinity, assuming the penetration depth to be small,
\begin{equation}
\sigma_x=\frac{1}{2}\mu_0\sigma \Re \int_0^\infty E_yH_z^* dx. \label{16}
\end{equation}
Here we have gone over to the complex  representation. In the metal $(x \geq 0)$, we have
\begin{equation}
E_y= \frac{kE_0}{\alpha}(1-i)e^{-\alpha x}\cdot e^{i(\alpha x -\omega t)},   \label{17}
\end{equation}
\begin{equation}
H_z=   \frac{kE_0}{\mu_0\omega}\left[ 2+(i-1)\frac{k}{\alpha}\right]e^{-\alpha x}\cdot e^{i(\alpha x-\omega t)}, \label{18}
\end{equation}
with $k=n\omega/c$ and
\begin{equation}
R=1-2k/\alpha, \quad \tan \delta=-k/\alpha. \label{19}
\end{equation}
Again, we see that there is no direct relationship to the photon momentum; only the Lorentz force on charge carriers is involved. This method of calculating $\sigma_x$ was used also by Peierls \cite{peierls76}, who derived the result (\ref{15}) in the limit $R=1$.

(3) Finally, the pressure can be found via the following argument, which  for us is the one of  primary interest. Assume that the electromagnetic energy-momentum tensor for the incident wave is divergence-free,  $\partial_\nu S_{\mu\nu}^{(i)}=0$. As all fields are functions of the argument $(x-ct/n)$ it follows that $(S_{xx}^{(i)}-cg_x^{(i)}/n)$ is a constant. This constant vanishes when the fields vanish, and so we conclude that  the relationship
\begin{equation}
S_{xx}^{(i)}=cg_x^{(i)}/n \label{20}
\end{equation}
 must hold. This means  that the momentum of the incident wave propagates with the velocity $c/n$. This expression can, in view of (\ref{11}), be rewritten in the Minkowski case as
\begin{equation}
S_{xx}^{{\rm M}(i)}=\frac{n}{c}S_x^{(i)}. \label{21}
\end{equation}
A similar term $nRS_x^{(i)}/c$ holds for the reflected wave, and so we are back to the expression (\ref{15}) for the pressure $\sigma_x$ on the wall.

We then arrive at the following general interpretation: The physical force density in the medium is  ${\bf f}^{\rm A}$. The Abraham term ${\bf f}^{\rm Aterm}$ in (\ref{14}) produces a mechanical momentum
\begin{equation}
{\bf g}_{\rm mech}^{\rm A}=\frac{n^2-1}{c^2}\bf E\times H, \label{22}
\end{equation}
running together with the field. The whole propagating momentum density is accordingly the Minkowski momentum,
\begin{equation}
{\bf g}^{\rm A}+{\bf g}^{\rm A}_{\rm mech}={\bf g}^{\rm M}. \label{23}
\end{equation}
This is in our opinion the chief message of the experiment of Jones {\it et al.}. The simple argument above shows the applicability of the Minkowski alternative in practical cases, and it shows the close relationship, rather than conflict, that exists between the two alternatives in the steady optical case. The divergence-free property of the Minkowski tensor implying, as mentioned earlier, the four-vector property of total energy and momentum,  is the main reason why it has turned out to be so convenient in  canonical, especially quantum-mechanical, contexts.

\subsection{Radiation pressure measured by the photon drag effect}

In a related experiment of Gibson {\it et al.} \cite{gibson80}, an ingenious use was made of the photon drag effect known from semiconductor physics \cite{gibson70} to isolate and measure the electromagnetic pressure. This appears to be the first time that the photon drag has been taken into consideration in connection with electromagnetic momentum.

Generally, the photon drag effect is is the generation in a rod of semiconductor material of a longitudinal electric field $E$ due to the transfer of momentum from radiation to the electrons in the valence or conduction bands. For the elemental semiconductors germanium and silicon used in the experiment the drag effect can be described in terms of two nonzero components which reduce, in the special case when the wave vector of incident radiation and the generated field $E$ are both collinear with a [100] crystal direction, to one single component. This component is equal to $E/I$ (for reasons of comparison with the mentioned papers we denote in the present subsection the Poynting vector by $I$). The field $E$ is a consequence of charges being driven down the rod. Under open-circuit conditions the current in the rod must be zero; the quantity measured in the experiment is the voltage between the two ends, and so $E$ can be determined. Problems arising from carrier back-diffusion can be avoided by making use of  short  radiation pulses ($<$ 200 ns). Theoretically, one can  determine  the value of $E$ by equating the electric force per carrier, $eE$ ($e$ being the elementary charge), to the rate of momentum taken away from the incident wave per particle. Let $\sigma_a$ be the   absorption  cross section for the carriers.  This quantity is known experimentally. If $p$ denotes the photon momentum in the medium, the product of $p$ and the factor $1/(\hbar \omega e)$ must be equal to $E/(I\sigma_a)$. This relationship can be written as
\begin{equation}
\frac{I \sigma_ap}{\hbar \omega}=eE. \label{24a}
\end{equation}
This relation requires that the wavelengths are long. Gibson {\it et al.} investigated the far-infrared region, up to a wavelength of 1.2 mm, and found that these long wavelengths the effect turned out to be independent of the semiconductor band structure and dependent only on the radiation pressure in the dielectric.

The authors examined different alternatives for the momentum $p$ in (\ref{24a}), and found only one alternative to agree with the observations, namely the Minkowski alternative $p^{\rm M}$.

Again, we  see the same kind of behavior as in the Jones experiment: experimentally the total momentum in the traveling wave is found to be  the Minkowski momentum. The theory of this experiment has also been considered before, in some detail, in Ref.~\cite{brevik86}.

\subsection{Measurement of photon recoil momentum in a Bose-Einstein condensate}

The last example we shall consider within the category of already existing experiments is the photon recoil experiment of Campbell {\it et al.} \cite{campbell05}, measuring the photon recoil momentum in a Bose-Einstein condensate (BEC).  The condensate was created in a so-called Ioffe-Pritchard magnetic trap. A BEC condensate has a much higher density than laser cooled atomic clouds, thus facilitating the observation of the refractive index-dependence of the recoil frequency. The elongated $^{87}$Rb condensate, containing $1.5\times 10^6$ atoms, had a Thomas-Fermi radius of $8~\mu$m in the radial direction and $90~\mu$m in the axial direction. The recoil frequency was measured by means of a two-pulse Ramsey interferometer using non-resonant laser light.

We do not go into any detail concerning this experiment, but point out that the recoil momentum of atoms caused by the absorption of a photon turned out to be $\hbar k=\hbar n\omega /c$, where $n$ is the refractive index of the gas. Once again we experience the same kind of behavior: the total propagating photon momentum in the medium is the Minkowski momentum.

\section{Two proposals for experiments}

We close this paper by presenting two ideas about experiments that could in principle be able to tell us more about the force and momentum balance in the radiation field in matter.

\subsection{Possibility of measuring the Abraham force using whispering gallery modes}

As mentioned above, in stationary high-frequency fields such as in optics the Abraham term ${\bf f}^{\rm Aterm}$   fluctuates out when averaged over a period. It might seem therefore that it is not possible to measure this force directly. It is in principle possible, however, to
circumvent this difficulty by making use of an {\it intensity modulated} optical beam whereby the time derivative in (\ref{14}) varies slowly. It is natural to think of  whispering gallery modes as a convenient experimental tool in this context. Such modes are conventionally produced in microspheres and possess a large circulating power, in excess of 100 W. Since  the field energy is concentrated along the rim,  the arrangement becomes convenient for measuring the azimuthal radiation-induced Abraham torque on the sphere when it is suspended as a pendulum in the gravitational field. The arm in the torque calculation is essentially the same as the radius of the sphere. Our proposal is simply  a generalization of the idea behind the experiment of Walker {\it et al.} \cite{walker75,walker75a}  (cf. also the discussion in \cite{brevik79}), to the case of optical frequencies. These authors showed the existence of the Abraham force at {\it low} frequencies, at which the oscillations themselves are detectable.

Our idea was analyzed in some detail in Ref.~\cite{brevik10}, where both a cylinder and a sphere were considered as candidates for the torsional pendulum. Here, let us only sketch some essentials of the theory for the cylinder, as it is the simplest alternative. Let the laser beam in the cylinder be modulated with a frequency $\omega_0$ much lower than optical frequencies, so that the  power $P$ and the corresponding Poynting vector $S_\phi$ are
 \begin{equation}
 P=P_0 \cos \omega_0 t, \quad S_\phi=S_0\cos \omega_0t. \label{24}
 \end{equation}
 The azimuthal Abraham force density $f_\phi^{\rm A}=f_\phi^{\rm Aterm}$, given by  (\ref{14}), becomes
 \begin{equation}
 f_\phi^{\rm A}=-\frac{n^2-1}{c^2}\omega_0S_0\sin \omega_0t, \label{25}
 \end{equation}
 giving by integration over the volume  the following torque $N_z^{\rm A}$ on the cylinder ($z$ is the vertical axis and $a$ is the radius)
 \begin{equation}
 N_z^{\rm A}=-\frac{n^2-1}{c^2}2\pi a^2\omega_0P_0\sin \omega_0t. \label{26}
 \end{equation}
 We   insert $a=100~\mu$m, $P_0=100$ W as typical parameters. Moreover,  in view of the very small buildup and ringdown times for this kind of oscillator, we may  assume the  large value $\omega_0=1000~{\rm s}^{-1}$ for the modulation frequency. Then,
 \begin{equation}
N_z^{\rm A} \sim 0.7\times 10^{-19}~\rm Nm. \label{27}
\end{equation}
This is small, even smaller than the torque $10^{-16}$ Nm in the classic Beth experiment measuring photon spin \cite{beth36}. To measure the Abraham torque  would seem difficult, but not impossible.

It is instructive here to look also at the angular deflection $\phi$ of the cylinder, instead of the magnitude of the torque. These aspects are discussed in \cite{brevik10}, and will be omitted here. Typical angular deflections at resonance, $\phi=\phi_{\rm max}$, are estimated to be of order $10^{-8}~\rm rad~s^{-1}$ or less.

 We finally point out that it is not necessary  to measure $N_z^{\rm A}$ very accurately in order to confirm, or at least support, the existence of the Abraham torque. According to Minkowski there should be no azimuthal force, and thus no torque, at all.

\subsection{Possibility of measuring microsphere movement driven by an optical pulse}

The final proposition to be discussed here (not discussed earlier in the literature) starts from the  radiation pressure  experiment reported recently  by Li {\it et al.} \cite{li11}.  A short laser light pulse of duration 10ns and energy $\mathcal{H}=5.9~\mu$J was emitted in the horizontal $x$ direction from a tapered fiber, the  tip diameter lying in the region from $1.9~\mu$m to $2.6~\mu$m. The pulse was impinging normally on a glass microsphere of  diameter  $2a=46~\mu$m, thus much greater than the transverse diameter of the pulse, meaning that the curvature of the sphere was unimportant for the magnitude of the radiation pressure. A striking property of this experiment is the large magnitude of the observed effect: the sphere was observed to move  a distance 298 $\mu$m during a period of 7 ms, before it came to rest. The maximum velocity, about 3 ms after the arrival of the pulse, was 8.1 cm/s. This velocity is much larger than that usually encountered in radiation pressure situations for microparticles; for instance, in  the Kawata-Sugiura  experiment \cite{kawata92} measuring the longitudinal driving of a microsphere in the evanescent field, typical velocities were only a few micrometers per second. The large magnitude of the movement in \cite{li11} makes this setup generally attractive for applications in microphysics.

Now, the explanation given in \cite{li11} focused on the radiation pressure as the sole factor responsible for the displacement. As was   later pointed out by Pozar and Mozina \cite{pozar12},  the radiation pressure in the experiment was far to small to account for the observed large velocity. Instead, they argued that another factor was important in this experiment, namely the mass recoil following a dielectric breakdown near or on the sphere of the microsphere. The maximum fluence of the laser pulse coming out of the fiber tip was 140 J/cm$^2$, which is about the same as the surface damage threshold for the glass material. It was therefore concluded that
the most probable explanation was  the ignition of a  plasma in the focused region. We find this explanation reasonable, and shall in the following denote the ablation-induced momentum transfer to the sphere by $\Delta G$. Its magnitude is theoretically unknown.

Let us consider the following idealized model: Let the microsphere of radius $a$ be suspended in the gravitational field, such that it is practically free to move after being hit by the laser pulse. Assume that the surrounding medium, a fluid, has dynamic viscosity $\mu$. The mechanical momentum transferred to the sphere is the quantity $\Delta G$ just mentioned, plus the photon momentum taken up from the incident field. For simplicity we assume that the latter is completely absorbed. Assuming Minkowski theory, this quantity is $n\mathcal{H}/c$, where $n$ is the refractive index of the fluid. Thus when the velocity is at maximum, the momentum balance reads
\begin{equation}
Mv_{\rm max}=\Delta G+n\mathcal{H}/c,    \label{28}
\end{equation}
$M$ being the mass  of the sphere. The movement becomes thereafter slowed down because of viscous drag $D$.  As the Reynolds number is small, we can use the Stokes formula $D=6\pi \mu av$. Solving the equation of motion we get
\begin{equation}
v=v_{\rm max}\exp(-6\pi \mu at/M), \label{29}
\end{equation}
taking $t=0$  to correspond to $v=v_{\rm max}$. By integrating from $t=0$ to $t=\infty$ we find the  total displacement $L$,
\begin{equation}
L=\frac{Mv_{\rm max}}{6\pi \mu a}=\frac{\Delta G+n\mathcal{H}/c}{6\pi \mu a}. \label{30}
\end{equation}
Repeating the experiment with another fluid surrounding the microsphere, the unknown quantity $\Delta G$ can be eliminated. This procedure assumes that the following assumption holds: $\Delta G$ depends on the {\it amount of absorbed energy} $\mathcal{H}$ {\it only}, and is independent of the nature of the fluid. The assumption appears natural, since ablation is a high-temperature phenomenon. We  assume the same value of $\mathcal{H}$ in the two cases.

Let for definiteness the first surrounding medium to be air (subscript zero), with $n_0=1, \mu_0=1.8\times 10^{-5} {\rm Pa~s}$, and the second medium a liquid with refractive index $n$ and viscosity $\mu$. The corresponding displacements are $L_0$ and $L$. Then,
\begin{equation}
\left(\frac{L}{L_0}\right)^{\rm M}=\frac{\mu_0}{\mu}\left[ 1+\frac{\mathcal{H}(n-1)}{6\pi acL_0\mu_0}\right], \label{31}
\end{equation}
where we have applied a superscript M to notify that this result is according to the Minkowski theory. If we instead assumed that it is the Abraham momentum which is the whole traveling momentum, we would have to replace the pulse-deposited momentum $n\mathcal{H}/c$  by  $\mathcal{H}/(nc)$. That would in turn imply
\begin{equation}
\left(\frac{L}{L_0}\right)^{\rm A}=\frac{\mu_0}{\mu}\left[ 1+\frac{\mathcal{H}}{6\pi acL_0\mu_0}\left(\frac{1}{n}-1\right)\right]. \label{32}
\end{equation}
The difference between these two predictions are very small, as we would expect. For definiteness we may adopt the value $\mathcal{H}=5.9~\mu$J from above, put $a=25~\mu$m, and $L_0=300~\mu$m. We then estimate that the second terms in (\ref{31}) or (\ref{32}) are of magnitude
\begin{equation}
\frac{\mathcal{H}}{6\pi acL_0\mu_0} \approx 7.7\times 10^{-3}. \label{33}
\end{equation}
As before, we find the difference between the Abraham and Minkowski predictions to be quite small, and most likely difficult to measure.

\section{Conclusion}

The natural choice for the electromagnetic momentum in practical problems is the Minkowski momentum, serving the total traveling momentum of an electromagnetic wave. All experiments in optics that we are aware of, are most conveniently described in terms of this quantity, and the great adaptability of the Minkowski tensor to the canonical formalism, in particular in a quantum-mechanical context, is worth noticing.

From a fundamental point of view it is however the Abraham force which turns out to be the natural choice. This force produces an accompanying mechanical momentum traveling together with the wave, so that the total  momentum becomes just  the Minkowski expression. We have discussed, in particular, the possibility of measuring the Abraham force in optics using an intensity modulated optical wave, in analogy to the low-frequency Lahoz-Walker experiment \cite{walker75,walker75a} (section 5.1), but the effect seems very difficult to measure in practice, at least when using conventional field strengths. The smallness of the difference between the Abraham and Minkowski predictions in optics appears to be quite general.

We ought to mention again that all electrostrictive forces have been omitted. Usually, they do not come into play in connection with electromagnetic momentum.

Finally, to put our comments above into a wider perspective, we note that there exist alternative approaches to the conventional Abraham-Minkowski approach which we have been following. Thus the papers of Peierls \cite{peierls76} and Einstein-Laub \cite{einstein08} ought to be mentioned in this context, as well as the recent paper of Ravndal \cite{ravndal12}. A closer analysis of some of the earlier papers in this category is given in \cite{brevik79}.


\end{document}